\documentclass[11pt,a4paper]{article}
\usepackage[utf8]{inputenc}
\usepackage[T1]{fontenc}
\usepackage{lmodern}
\usepackage{microtype}
\usepackage[margin=2.5cm]{geometry}
\usepackage{setspace}
\usepackage{parskip}
\usepackage{graphicx}
\usepackage{booktabs}
\usepackage{enumitem}
\usepackage{hyperref}
\hypersetup{colorlinks=true,linkcolor=blue,citecolor=blue,urlcolor=blue}
\usepackage{titling}
\usepackage{fancyhdr}
\usepackage{caption}
\usepackage{amsmath,amssymb}
\usepackage{xcolor}
\usepackage{multicol}

\usepackage[style=numeric,
    maxbibnames=1,    
    giveninits=true,  
    sorting=none,     
    doi=false,
    url=false,
    isbn=false,
    eprint=false
    ]{biblatex}

\AtEveryBibitem{%
  \clearfield{title}
  \clearfield{pages}%
  \clearfield{publisher}%
  \clearfield{location}%
  \clearfield{url}%
  \clearfield{doi}%
  \clearfield{eprint}%
  \clearfield{month}%
}

\pretitle{\vspace*{-1cm}\begin{center}\LARGE\bfseries}
\posttitle{\par\end{center}\vskip 0.5em}
\preauthor{\begin{center}\large}
\postauthor{\end{center}}
\predate{\begin{center}\large}
\postdate{\end{center}}

\begin{document}

\begin{titlepage}
  \centering
  \vspace*{1cm}
  {\Huge\bfseries Expanding Horizons \\[6pt] \Large Transforming Astronomy in the 2040s \par}
  \vspace{1.5cm}

  { \LARGE{\textbf{Super-Eddington Accretion through a Multiwavelength Lens}} \\
  \Large{\textbf{Searching for Counterparts of Ultraluminous X-ray Sources}}\par}
  \vspace{1cm}

  \begin{tabular}{p{4.5cm}p{10cm}}
    \textbf{Scientific Categories:} & Compact objects, Accretion, Time-domain \\
    \\
    \textbf{Submitting Author:} & Name: Roberta Amato\\
    & Affiliation: INAF-OAR (IT) \\
    & Email: roberta.amato@inaf.it\\
    \\
    \textbf{Contributing authors:} & Matteo Bachetti (INAF-OAC, matteo.bachetti@inaf.it)\\ & Roberto Soria (INAF-OAT, roberto.soria@inaf.it)\\ & Andr\'es G\'urpide (University of Southampton, a.gurpide-lasheras@soton.ac.uk)\\ & Matteo Imbrogno (ICE-CSIC, mimbrogno@ice.csic.es)\\ & Chiara Salvaggio (INAF-OAB, chiara.salvaggio@inaf.it)\\ & Ruben Salvaterra (INAF-IASF MI, ruben.salvaterra@inaf.it)\\ & Melania Del Santo (INAF-IASF PA, melania.delsanto@inaf.it) \\ & Simone Scaringi (Durham University,  simone.scaringi@durham.ac.uk)\\ & Piergiorgio Casella (INAF-OAR, piergiorgio.casella@inaf.it)\\
    & Anna Wolter (INAF-OAB, anna.wolter@inaf.it)
  \end{tabular}




\end{titlepage}


\section{Introduction and Background}
\label{sec:intro}

Super-Eddington accretion constitutes one of the most challenging frontiers in contemporary astrophysics, manifesting across a wide range of phenomena, from the rapid growth of early supermassive black holes (SMBHs) to accretion onto nearby, stellar-mass compact objects, and tidal disruption events. Recent surveys with facilities such as JWST have uncovered quasars with billion-solar-mass black holes at $z>6$. Accounting for the formation of such massive objects requires prolonged phases of accretion likely exceeding the canonical Eddington limit, yet the physical processes enabling such growth remain poorly understood.

Fortunately, the nearby Universe provides accessible laboratories where super-Eddington accretion can be studied in detail: Ultraluminous X-ray sources (ULXs). These are X-ray binaries subject to super-Eddington mass transfer rates, identified in X-ray observations based on their off-nuclear position in nearby galaxies and their X-ray luminosities exceeding the Eddington limit for a BH of 10 $M_\odot$ ($L_\mathrm{Edd} = 10^{39}$ erg s$^{-1}$; see reviews \cite{Kaaret2017, King2023_review,PintoWalton2023_review_arxiv}). Despite being known since the 80s, only in the last decade X-ray coherent pulsations have been discovered for the first time in a few ULXs (generally referred to as pulsating ULXs; PULXs), with periodicities ranging from $\approx$0.4\,s to tens of seconds \cite{Israel2025}, ultimately proving the nature of the compact objects as neutron stars (NSs). ULXs are now understood to be powered by accretion onto compact objects, with masses spanning $\sim$1--$10^5$\,$M_\odot$, encompassing NSs, stellar-mass and intermediate-mass BHs. 

Detecting pulsations from ULXs is intrinsically difficult, as their large distances ($\gtrsim10$\,Mpc) demand long exposures (of the order of 100\,ks with current X-ray facilities) and the pulse signal is often suppressed or distorted by extreme spin derivatives ($\dot{P}\sim10^{-10}$\,s\,s$^{-1}$), strong orbital Doppler modulations, complex geometrical configurations, involving thick discs and relativistic disc outflows, and on-/off-states. Consequently, the paucity of confirmed PULXs is unsurprising, and a substantially larger fraction of ULXs may harbour NSs whose pulsations remain undetected.

Current observational facilities do not permit conclusive pulsation searches from ULXs at optical, UV, or IR wavelengths, primarily due to the source's distances. Optical pulsations have been observed only in closer X-ray pulsars, such as transitional millisecond pulsars \cite{Ambrosino2017}, tentatively in GX 1+4, 4U 1626–67 \cite{Chakrabarty2001,Chakrabarty1997}, and in Her X-1 \cite{Middleditch1985}, where they are generally attributed to reprocessing of the X-ray pulsations from the accretion stream, disc, or the companion star’s surface. By analogy, optical or UV pulsations could in principle arise in PULXs through reprocessing of the X-ray radiation in the disc outflows photosphere, or in the donor star atmosphere. In addition,  several ULXs, including pulsating ones, exhibit mHz quasi-periodic oscillations (QPOs) interpreted as signatures of precessing inner accretion discs \cite{Veresvarka2025}. Suitably designed future facilities could enable the search for such QPOs also at other wavelengths, while simultaneous optical and X-ray observations could allow the search of time lags between the two wavebands, providing valuable constrains on the system geometry and spatial origin of the reprocessed optical emission.

The mass-accretion rate is one of the main parameters dictating whether a system can enter the super-Eddington regime. This rate strongly depends on the nature of the donor star, making the identification of the companion crucial in determining whether the system is genuinely accreting at super-Eddington rates or instead hosts a particularly massive compact object. Identifying the companion star in ULXs is challenging. Their extragalactic distances can result in faint counterparts ($M_V \sim -6$ for the nearby systems \cite{Tao2011}) going undetected and multiple sources lying within the X-ray positional uncertainty. 
Furthermore, emission from the accretion disc or the wind photosphere can contaminate, or even dominate, the observed photometric and spectroscopic signals \cite{Grise2012}. Searches with optical and IR ground-based and space facilities have identified counterparts for a few nearby ULXs \cite{Heida2014, Heida2016,Lau2019, Lopez2020, Allak2024, Tao2011, Gladstone2013}, including two pulsating ULXs \cite{Motch2011,Heida2019}, against $\sim$2000 ULXs detected in the X-rays so far \cite{Tranin2024_cat,Walton2022_ULXcat}. 
The securely identified companion stars are mostly giants (O, B type, e.g. \cite{Motch2011}) or (red) supergiants (e.g. \cite{Heida2014}), but there remains discussion as to whether the blue appearance of the counterparts is being skewed by the disc emission. Only in a few cases, the (blue) optical emission is firmly attributed to the irradiated accretion disc \cite{Soria2012}.

Optical spectroscopy and photometry can also provide the orbital parameters of the system, enabling constraints on the mass of the compact object and the evolution of the binary. 
Optical observations can also be used to determine the ULX association with young star clusters, star-forming regions, or reveal the presence of ionised nebulae extending hundred of parsecs, produced by the interaction of the disc outflows with the interstellar medium. All combined, these diagnostic give information about the age of the system, the type of progenitor, and the duration of the super-Eddington accretion phase \cite{Gurpide2022_MUSE}.

The evolutionary history of ULXs is highly relevant for binary evolution studies, as binary synthesis population models often involve high-mass stars and extreme mass ratios to form a ULX-type system. Consequently, ULXs are prime candidate progenitors for compact-object binaries that will eventually merge and produce gravitational waves (GW), detectable by present and future generations of GW detectors (LIGO, Virgo, KAGRA, ET, etc.) \cite{Mondal2020}. For instance, ULXs may reside in compact binaries with a Wolf-Rayet companion. These systems are characterised by tight orbital configurations ($P_{\rm orb} \lesssim 10$\,d) and rapid evolutionary timescales, and indeed considered promising progenitors of GW merger events. Accurate identification of compact binaries requires precise determination of the orbital period, achievable through high-cadence optical photometry and phase-resolved spectroscopy on timescales of less than one hour. Representative examples of nearby, compact, high-mass X-ray binaries include Cyg X‑3 \cite{Zdziarski2013,Korb2025} and Cir X‑1 \cite{Qiu2019}.

The determination of the nature of the compact object in ULXs may help alleviate the tension between GW- and electromagnetically-detected BHs. In particular, the mass distribution of ground-based GW-detected BHs has a long tail extending to masses $\sim$100\,$M_\odot$, while BHs in X-ray binaries for which a dynamical mass is available only span the $\sim$5--30\,$M_\odot$ range. Some non-pulsating, extremely luminous, ULXs could harbour the missing, heavier BHs ($\sim$30--100\,$M_\odot$), bridging the gap between the two distributions.

While super-Eddington accretion in ULXs is sustained over long timescales ($\sim10^5$\,yr), transient episodes of super-Eddington accretion are also observed, for example, in tidal disruption events (TDEs) and their stellar-mass analogues, micro-TDEs. The latter are caused by the full or partial disruption of stars or planets captured by stellar-mass compact objects (BHs or NSs), with the subsequent formation of a debris accretion disc. Compared to classical TDEs from SMBHs, micro-TDEs evolve on much shorter timescales (10$^3$--10$^4$\,s), but could nonetheless manifest as high-energy or multi-wavelength transients \cite{Perets2016, Xin2024}, as recently proposed for GRB 250702B \cite{Carney2025,OConnor2025}. In this case, fast optical photometry and spectroscopy can help tracking the rise and decay of the optical emission, constraining debris fallback and accretion timescales, as well as dimensions and temperatures of the system.

\section{Open Science Questions in the 2040s}
\label{sec:openquestions}

\begin{enumerate}[nosep]
    \item \textbf{Can we detect optical pulsations, QPOs, time lags?} Optical pulsations might come from the illumination of the outflows or the companion star's surface from the pulsar. The detection of optical pulsations would provide an additional venue for pulse detection in case the alignment of the spin axis is unfavourable for X-ray pulse detection. 
    If both optical and X-ray pulsations were detected, measuring a time lag between them could help constrain the geometry of the system (e.g. measure the distance between the NS and the illuminated medium). The study of aperiodic and quasi-periodic variability, whose presence in ULXs is well known, as well as pulse-phase spectroscopy, could help shed light on the details of the accretion physics at super-Eddington rates.
    \item \textbf{Which are the companion star types in ULXs? Can we disentangle the contribution from the accretion disc and the stellar emission?} While the companions have been identified in some cases, a statistically significant sample is still lacking to compare observations to predictions from binary population synthesis studies. Disentangling the companion emission from that of the accretion disc can provides critical constraints on the mass-transfer rate, which in turn determines the accretion rate onto the compact object. This could help constrain the compact-object mass, offering crucial insight into the evolutionary pathways of these binaries and their environment, including the duration of super-Eddington phases and the overall population properties.
    \item \textbf{Can we fill the gap on the mass distribution of compact objects?} The most luminous ULXs remain strong candidates for intermediate-mass BHs, potentially bridging the gap between stellar-mass BHs and the more massive BHs detected via GW mergers. Dynamical assessments of these systems, for instance through high-cadence optical photometry and phase-resolved spectroscopy on timescales of less than one hour, could provide insightful constraints on the mass of the compact objects.
    \item \textbf{How much energy is released during the super-Eddington phase? How long does it last?} Large-collecting area optical or IR observations probing high-excitation lines can reveal the faint (H$\alpha$$\sim10^{-16}$\,erg\,cm$^{-2}$\,arcsec$^2$ for nearby $D\lesssim10$\,Mpc objects) shock-ionised bubbles around ULXs produced by the accretion disc winds. These can be used to determine the lifetime of the super-Eddington phase as well as quantify the amount of mechanical energy injected in the environment.
\end{enumerate}

\section{Technology and Data Handling Requirements}
\label{sec:tech}

We advocate for one or more facilities that would help complement the current knowledge on ULX science with fast photometry or spectroscopy in the optical, UV and IR wavebands. Such facilities should have a large collecting area to gather sufficient photon counts for fast timing on faint extragalactic sources, and time resolution of seconds or less, to recover pulsations in the Hz regime. The facilities should also work in synergy with present and upcoming missions, especially NewAthena, which will substantially enhance the number of detected ULXs in the X-ray, and future GW detectors. 


\begin{multicols}{2}
\printbibliography
\end{multicols}

\end{document}